\documentclass[
    twocolumn,
    superscriptaddress,
    amsmath,amstex,amssymb,
    aps, pra,
    floatfix,
    longbibliography,
    showkeys,
    letterpaper,
    reprint
    ]{revtex4-2}

\usepackage{graphicx}
\usepackage{dcolumn}%
\usepackage{bm}
\usepackage{lipsum}
\usepackage{gensymb}
\usepackage{physics}
\usepackage{booktabs}
\usepackage{amsmath,amssymb,amsfonts}
\usepackage{algorithmic}
\usepackage{textcomp}
\usepackage{enumitem}
\usepackage{soul}
\usepackage{xspace}

\usepackage{multirow}
\usepackage[usenames,dvipsnames,table,xcdraw]{xcolor}
\def\BibTeX{{\rm B\kern-.05em{\sc i\kern-.025em b}\kern-.08em
    T\kern-.1667em\lower.7ex\hbox{E}\kern-.125emX}}

\usepackage{hyperref}
\hypersetup{
  colorlinks   = true, 
  urlcolor     = blue, 
  linkcolor    = red, 
  citecolor   = red 
}

\newcommand{\hlsfml}{\texttt{hls4ml}\xspace}

\newcommand{\reportnumber}{FERMILAB-PUB-24-0925-ETD-PPD}

\usepackage[angle=0,scale=1,color=black,firstpage=true,opacity=1]{background}

\SetBgContents{\small\texttt{\reportnumber}}
\SetBgPosition{current page.south east}
\SetBgHshift{-4cm}
\SetBgVshift{1cm}

\begin{document}

\title{End-to-end workflow for machine learning-based qubit readout with QICK and hls4ml}

\author{Giuseppe~Di~Guglielmo}
    \email{gdg@fnal.gov}
    \affiliation{Fermi National Accelerator Laboratory, Batavia, IL 60510, USA}
\author{Botao~Du}
    \affiliation{Department of Physics and Astronomy, Purdue University, West Lafayette, IN 47907, USA}
\author{Javier~Campos}
    \affiliation{Fermi National Accelerator Laboratory, Batavia, IL 60510, USA}
\author{Alexandra~Boltasseva}
    \affiliation{Elmore Family School of Electrical and Computer Engineering, Birck Nanotechnology Center and Purdue Quantum Science and Engineering Institute, Purdue University, West Lafayette, IN 47907, USA}
\author{Akash~V.~Dixit}
    \affiliation{University of Chicago, Chicago, IL 60637, USA}
\author{Farah~Fahim}
    \affiliation{Fermi National Accelerator Laboratory, Batavia, IL 60510, USA}
\author{Zhaxylyk~Kudyshev}
    \affiliation{Elmore Family School of Electrical and Computer Engineering, Birck Nanotechnology Center and Purdue Quantum Science and Engineering Institute, Purdue University, West Lafayette, IN 47907, USA}
\author{Santiago~Lopez}
    \affiliation{Department of Physics and Astronomy, Purdue University, West Lafayette, IN 47907, USA}
\author{Ruichao~Ma}
    \affiliation{Department of Physics and Astronomy, Purdue University, West Lafayette, IN 47907, USA}
\author{Gabriel~N.~Perdue}
    \affiliation{Fermi National Accelerator Laboratory, Batavia, IL 60510, USA}
\author{Nhan~Tran}
    \affiliation{Fermi National Accelerator Laboratory, Batavia, IL 60510, USA}
\author{Omer~Yesilyurt}
    \affiliation{Elmore Family School of Electrical and Computer Engineering, Birck Nanotechnology Center and Purdue Quantum Science and Engineering Institute, Purdue University, West Lafayette, IN 47907, USA}
\author{Daniel~Bowring}
    \affiliation{Fermi National Accelerator Laboratory, Batavia, IL 60510, USA}
    
\date{\today}

\begin{abstract}
We present an end-to-end workflow for superconducting qubit readout that embeds co-designed Neural Networks (NNs) into the Quantum Instrumentation Control Kit (QICK). Capitalizing on the custom firmware and software of the QICK platform, which is built on Xilinx RFSoC FPGAs, we aim to leverage machine learning (ML) to address critical challenges in qubit readout accuracy and scalability. The workflow utilizes the \texttt{hls4ml} package and employs quantization-aware training to translate ML models into hardware-efficient FPGA implementations via user-friendly Python APIs. We experimentally demonstrate the design, optimization, and integration of an ML algorithm for single transmon qubit readout, achieving 96\% single-shot fidelity with a latency of 32\,ns and less than 16\% FPGA look-up table resource utilization. Our results offer the community an accessible workflow to advance ML-driven readout and adaptive control in quantum information processing applications.
\end{abstract}

\keywords{Quantum Computing; Superconducting Qubit Readout; Machine Learning; Radio Frequency System-on-Chip (RFSoC)}

\maketitle

\section{Introduction}
\label{sec:intro}

Quantum technologies hold the promise to transform a range of applications from computation and communication to sensing.
Realizing these quantum advantages, however, requires scaling from current small-scale prototypes to large-scale quantum processors with increasingly complex qubit arrays.
As the number of qubits grows, so does the need for a commensurately scalable and efficient classical co-processing infrastructure. The software, firmware, and hardware responsible for controlling and reading out the quantum states must not only support the expanding qubit counts but also maintain the high fidelity and low latency critical for successful quantum operations.

Superconducting (SC) qubits have emerged as a leading platform for building large-scale quantum processors \cite{Kjaergaard2020}.
The rapid progress is thanks to the scalability of the lithographically fabricated superconducting circuits, the instrumentation ecosystem available at the operating frequencies of SC qubits, and advances in cryogenic systems to reach the millikelvin temperatures required for their quantum operation. Noisy Intermediate-Scale Quantum (NISQ) devices with more than a hundred qubits are now in operation, showing significant progress towards computational quantum advantages \cite{Kim2023-lv,QECbelowThresh2024, Gao2024-yc}. As these devices increase in complexity, there is a growing need for integrated control and readout solutions that combine functionality, flexibility, ease of deployment, and cost-effectiveness.

Recent developments in Radio Frequency System-on-Chip (RFSoC) technologies are accelerating this effort \cite{qick, Xu2021-QubiC, Tholn2022-Presto, Park2022-ICARUS}. 
By combining the capabilities of Field-Programmable Gate Arrays (FPGAs) with RF data converters, RFSoC-based systems provide real-time signal generation via direct digital synthesis in the microwave domain, efficient readout via fast analog-to-digital conversion, and low-latency signal processing.
In particular, the Quantum Instrumentation Control Kit (QICK) \cite{qick, Ding2024-QICK2} has seen growing adoption among laboratories developing superconducting qubits, as well as other experimental platforms.

\begin{figure}[t]
    \centering
    \includegraphics[width=0.9\linewidth]{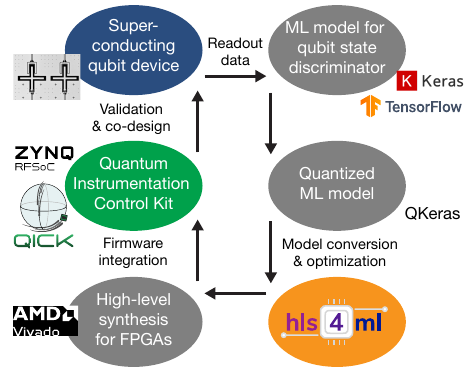}
    \caption{Overview of the QICK and \hlsfml workflow.
    }
    \label{fig:workflow-overview}
\end{figure}

In parallel to this hardware development, there is a crucial need to integrate the classical and quantum hardware with a robust and scalable software stack.
Software challenges include device calibration and tune-up, multiplexed readout, and fast adaptive control.
These challenges span both classical and quantum control in the presence of noise, crosstalk, and other errors.
For future fault-tolerant quantum machines, fast, high-fidelity measurements across large systems and real-time, low-latency adaptive feedback control are essential \cite{beverland2022assessingrequirementsscalepractical,mohseni2024buildquantumsupercomputerscaling}.

The rise of machine learning (ML) tools in classic computing provides a powerful toolset to develop adaptive, heuristic algorithms for qubit readout and control.
In particular, ML can be used in readout to account for effects that are difficult to compute analytically, such as non-linear behaviors in the time-evolution of signal traces including noise, multi-qubit correlations and crosstalk, and the evolution of the system dynamics over time due to external effects~\cite{gautam2024low,Duan2021-zg,PhysRevLett.114.200501,Maurya2023scaling,vora2024mlpoweredfpgabasedrealtimequantum}.
Ultimately, an active learning ML approach can provide a path to high-fidelity autonomous and adaptive qubit readout control systems.
Toward this goal, it is important to develop a platform for researchers to study, train, and deploy ML algorithms, bringing together expertise in quantum systems, machine learning, and readout electronics that meet the constraints of state-of-the-art experiments.  

In this paper, we present an open-source workflow based on QICK and \hlsfml to advance ML-based quantum control. 
The open-source \hlsfml package~\cite{Duarte:2018ite,fastml_hls4ml} enables users to translate ML models into low-level logic descriptions using High-Level Synthesis tools through accessible Python APIs. This allows for algorithm and experiment design without requiring detailed firmware development knowledge for FPGAs. 
Specifically, hls4ml supports the optimization of neural network designs in hardware with techniques such as quantization-aware training~\cite{jacob2017quantizationtrainingneuralnetworks} and pruning~\cite{cheng2024survey}, which minimize hardware resources and latency by tuning high-level parameters.

We demonstrate an ML-based qubit readout algorithm on QICK, integrating a neural network designed through \hlsfml.
This includes developing and deploying an end-to-end workflow for ML-based superconducting qubit readout and providing open-source tools for others to use in their research.
Our initial demonstration focuses on the readout of a single superconducting transmon qubit.
The ML implementation achieves a single-shot readout fidelity of 96\% with a latency of 32\,ns and the maximum FPGA resource usage of Look-Up Tables (LUTs) of 16\%.
The open-source end-to-end workflow for designing and integrating NN models into QICK is illustrated in Figure~\ref{fig:workflow-overview}. The code is made publicly available at Ref.~\cite{github-ml-quantum-readout}.

This paper is organized as follows. In Section 2, we describe the setup for the superconducting qubit system under study, visualize the readout data, and describe non-ML-based readout methods.  In Section 3, we detail the NN model training and optimization with \hlsfml for the available resources within the QICK readout system.  Section 4 describes the integration of the \hlsfml neural network intellectual property (IP) block with the QICK firmware and hardware.  We conclude in Section 5, describing the results of this study and what future capabilities this will enable.

\section{Superconducting Qubit Readout}
\label{sec:setup}

\subsection{Background}

In superconducting (SC) qubits, quantum information is encoded as excitations of electromagnetic resonances formed by SC circuit elements. The transmon qubit \cite{Koch2007-cz}, a workhorse in contemporary SC quantum processors, can be described as an anharmonic microwave resonator, where the SC Josephson junction provides the anharmonicity. The computational basis of the transmon consists of its two lowest energy levels, denoted as $\ket{0}$ and $\ket{1}$, which correspond to zero and one excitation (microwave photon) in the transmon, respectively. The anharmonicity of the transmon ensures unequal level spacing, enabling coherent control within the computational subspace using resonant microwave pulses.

The quantum state of a transmon qubit is measured by coupling the transmon to a linear microwave resonator \cite{Krantz2019-zv, PRXQuantum.2.040202}. The dispersive interaction between the readout resonator and the qubit induces a frequency shift in the resonator that depends on the qubit state. Consequently, the qubit states can be distinguished by probing the readout resonator at a frequency near its resonance and monitoring the transmitted or reflected quadrature signals. In a typical projective measurement, the quantum state collapses into one of the basis states, and the readout produces a single binary outcome.

Scaling SC quantum processors toward fault-tolerant operation and quantum error correction requires high-fidelity single-shot readout with minimal measurement duration. Achieving this fidelity depends on several design considerations that influence the distinguishability of the qubit states $\ket{0}$ and $\ket{1}$ \cite{Krantz2019-zv}. These factors include the linewidth of the readout resonator, the magnitude of the dispersive shift, noise from downstream amplifiers, and the number of photons collected during the measurement.
The latter is determined by the integration time and the readout power.

\begin{figure}[t]
    \centering
    \includegraphics[width=.95\columnwidth]{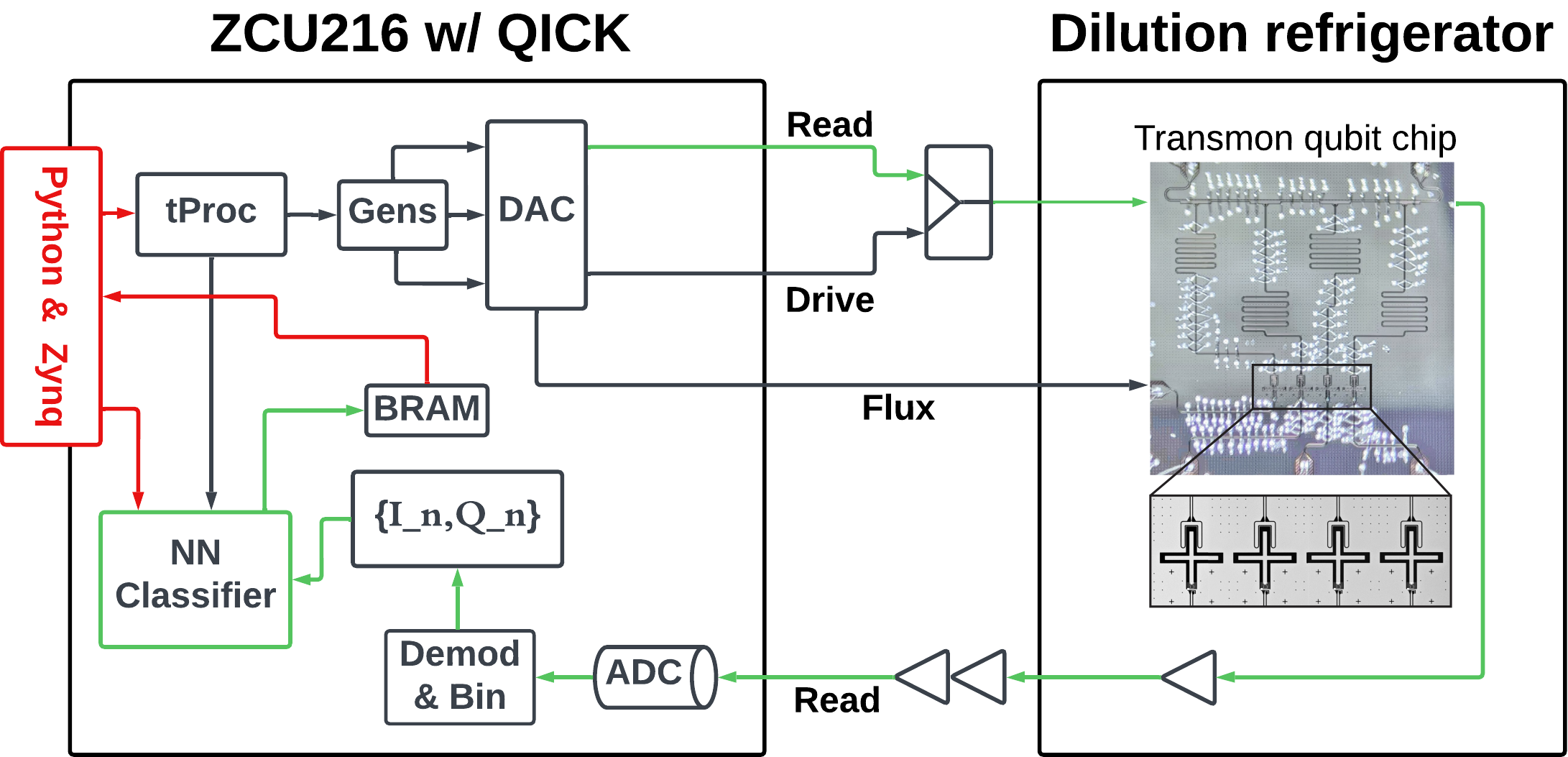}
    \caption{Experimental setup. The Xilinx RFSoC ZCU216 runs on custom QICK firmware and software for qubit control and readout. The qubit readout and drive pulses are combined before entering the fridge; the flux pulse is used for qubit cooling. The readout signal is amplified before being digitalized by the ADC.}
    \label{fig:exp-hardware-final}
\end{figure}

\subsection{Control and readout hardware}

The experiments in this work are conducted on a frequency tunable multi-transmon device, as described in \cite{du2024probing}. The hardware setup for qubit control and readout is illustrated in Fig.~\ref{fig:exp-hardware-final}.
The ML-based qubit readout is performed on a single qubit of frequency $\omega_q=2\pi \times 4.50$\,GHz, with all other qubits far-detuned in frequency and effectively decoupled. The corresponding readout resonator has frequency $\omega_r = 2\pi \times 6.32$\,GHz, and linewidth $\kappa_r = 2\pi \times 1.5$\,MHz from its coupling to the readout transmission line. The qubit-resonator coupling is $g = 2\pi \times 65$\,MHz, giving a qubit-state-dependent dispersive shift of $2\chi = 0.4$\,MHz. At the operating frequency, the transmon qubit has a relaxation time of $T_1 = 32\,\mu$s, and a dephasing time $T_2^* = 1.7\,\mu$s.

The Xilinx ZCU216 RFSoC evaluation board is running a modified version of the QICK firmware with an integrated ML classifier.
We utilize a mixer-free setup without any external local oscillators \cite{Ding2024-QICK2}:
The microwave pulses for qubit control and readout are directly synthesized with the ZCU216's 6.88\,GS/s digital to analog converters (DACs). rly, the readout signal from the SC qubit device is directly digitized by the ZCU216's analog to digital converters (ADCs) running at 6.88\,GS/s and demodulated on the FPGA to generate the time-series quadrature signals $I(t)$ and $Q(t)$.

\subsection{Single-shot qubit readout}

To characterize the readout fidelity, we first prepare the transmon qubit in the computational basis of ground state $\ket{0}$ and excited state $\ket{1}$. 

In our device, the transmon qubit has an excited state population of approximately 6\% at thermal equilibrium due to coupling to environmental noise. To reduce state preparation error from the thermal population, we perform a cooling step at the beginning of each experiment which utilizes the readout resonator as a dissipative cold reservoir. By applying an AC signal of frequency $\omega_r-\omega_q$ on the flux control line to modulate the transmon frequency, the transmon parametrically couples to the resonator and relaxes towards a lower excited state population \cite{du2024probing}.
Here, we use a flux modulation pulse of duration 2.3 $\mu$s, and an amplitude that modulates the qubit frequency by approximately $\pm 36$\,MHz. This corresponds to an effective resonant qubit-resonator coupling of $g_\text{eff} = 2\pi \times 0.65$\,MHz. The modulation pulse, with a frequency at the qubit-resonator detuning of approximately 1.8\,GHz, is generated using a separate DAC channel on the QICK-controlled ZCU216. With the cooling sequence, we prepare the qubit ground state with less than 1.6\% thermal population, limited by the thermal population in the readout resonator.repare the excited state, we first perform the cooling step and then apply a resonant microwave $\pi$-pulse at the qubit frequency with a Gaussian width $\sigma = 46$\,ns and total length $4\sigma$. The $\pi$-pulse error is $\leq 0.4$\%, characterized by measuring qubit population after applying repeated $\pi$-pulses.

Immediately following the initial state preparation, we send a 2.5\,$\mu$s square-shaped readout pulse to the readout resonator. After interacting with the coupled resonator-qubit system, the readout pulse returned to the ZCU216 and is digitized by the ADC to a discrete-time series of quadrature values $\{I_n,Q_n\}$, where $n$ labels the discrete time steps.
In our QICK setup, the readout pulse length of 2.5\,$\mu$s corresponds to $770$ ADC clock cycles, i.e. $n=\{1,...,770\}$. 
The experiment is repeated with a cycling period of 100 $\mu$s to collect statistics.
For the labeled training data, we used 500,000 shots each for qubits prepared in $\ket{0}$ and $\ket{1}$.
In the standard QICK firmware, the time series can either be saved in the on-board memory for offline processing, or time-averaged in real-time and saved as average values for each readout $\{\bar{I}, \bar{Q}\}$.
Additionally, the time-averaged values can be compared to a pre-configured threshold in real-time on the FPGA to yield a binary output and be used as logic input for conditional qubit control.
In our current implementation of ML-based readout, the time-series $\{I_n, Q_n\}$ is streamed to the NN classifier block on the FPGA and the NN prediction outputs are saved in the programmable logic block
memories (BRAM), as shown in Fig.~\ref{fig:exp-hardware-final}.

\begin{figure}
    \centering
    \includegraphics[width=.9\columnwidth]{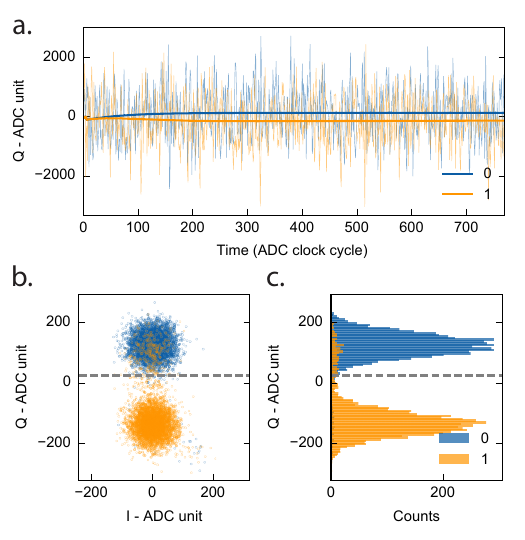}
    \caption{Typical single shot readout signal. The readout phase is adjusted to optimize the signal in \textit{Q} quadrature. (a) Single-shot readout trajectory and averaged readout trajectory. (b)(c) Time-averaged single-shot readout signal in the I-Q plane and its histogram (showing 5000 shots for better visibility). The gray dashed line represents the readout threshold for optimized readout fidelity.}
    \label{fig:exp_2}
\end{figure}

\subsection{Readout Fidelity from threshold methods}

Figure~\ref{fig:exp_2}(a) shows representative single-shot readout data and the time-series I/Q trajectories averaged over all training shots. The trajectories for states $\ket{0}$ and $\ket{1}$ separate within the first 100 clock cycles (approximately 325 ns), limited by the linewidth of the resonator. Beyond this initial period, the photon number in the resonator stabilizes, resulting in a near-constant separation between the average $\ket{0}$ and $\ket{1}$ trajectories.

We plot the time-averaged readout signal in the I/Q plane in Fig.~\ref{fig:exp_2}(b), revealing two distinct Gaussian-like distributions. A simple threshold can be applied to the time-averaged I/Q signal to predict the qubit state and quantify readout fidelity. By projecting the I/Q signal along the axis connecting the average values of the $\ket{0}$ and $\ket{1}$ state distributions, we generate the histogram shown in Fig.~\ref{fig:exp_2}(c), with the optimal threshold indicated by the gray dashed line.

The single-shot readout fidelity is defined as $\mathcal{F} \equiv 1 - \frac{1}{2}(P(0\vert 1) + P(1\vert 0))$. Here, $P(0\vert 1)$ is the probability of a qubit prepared in the excited state being misclassified as the ground state, and $P(1\vert 0)$ is the probability of the reverse classification error.

Using the full readout data (770 clock cycles), the thresholding (TH) method yields a single-shot fidelity of $\mathcal{F}=95.80 \pm 0.03$\%, with $P(0\vert 1)\approx 6.05 \pm 0.04$\% and $P(1\vert 0)\approx 2.34 \pm 0.02$\%.
Error bars represent the standard error of the mean, based on the training dataset size. 
The distinguishability of the qubit states is initially limited by the resonator bandwidth, and diminishes at longer times due to qubit relaxation. Therefore, a weighted time average of the I/Q signal can enhance performance when the threshold is applied. Using the optimal weights given by the matched-filter (MF) method \cite{PhysRevA.91.022118} with the full readout window, we obtain $\mathcal{F}=95.76 \pm 0.03 $\%, comparable to the simple thresholding (TH) method.

The observed single-shot infidelity results from several factors, including residual thermal population, qubit relaxation, and signal-to-noise ratio (SNR) of the amplifier chain.
Residual thermal population ($\approx 1.6$\%) contributes to errors in both $P(0\vert 1)$ and $P(1\vert 0)$. 
Additionally, qubit relaxation during the readout window ($T_1$ decay) accounts for $\approx 4$\% contribution to $P(0\vert 1)$.
These effects are evident in the histogram as asymmetric tails of the Gaussian distributions. 
The remaining infidelity ($<1$\%) arises from finite SNR in the readout amplifier chain and potential readout-induced qubit transitions not captured by $T_1$ decay.
Higher readout fidelity can be achieved by employing quantum-limited amplifiers to enhance SNR \cite{Macklin2015twpa}, incorporating Purcell filters to reduce qubit relaxation and allow faster readout \cite{PhysRevLett.112.190504, PhysRevApplied.10.034040}, and optimizing readout pulse shaping to minimize state-changing errors and readout duration \cite{PhysRevApplied.5.011001}.

To utilize the data more efficiently and reduce the size of the neural network (NN) classifier, we analyze the readout fidelity with truncated single-shot data, limiting the readout window to 400 clock cycles starting at different locations within the readout period, as shown in Fig.\,\ref{fig:exp_3}. Using the threshold (TH) method, we achieve an optimized fidelity of $\mathcal{F}=96.04 \pm 0.03$\% with $P(0\vert 1)\approx 5.56 \pm 0.04$\% and $P(1\vert 0)\approx 2.35 \pm0.02$\% when truncating the window to [100,500] clock cycles. The matched-filter (MF) method yields a comparable fidelity of $\mathcal{F}=96.01 \pm 0.03$\% for the same truncation window. A detailed discussion of truncation window optimization for the NN method is provided in Sec.\ref{subsec:design_space}.

\section{Neural Network Model Co-design}
\label{sec:design}
\subsection{Design strategy}

The primary objective of the neural network model is to facilitate efficient and accurate qubit readout within the firmware design pipeline.
Recently, various neural network architectures have found direct applications with transmon qubit systems.
Recurrent neural networks (RNNs), designed for sequential data, have been utilized to infer individual quantum trajectories of a superconducting qubit's evolution~\cite{PhysRevX.10.011006, PhysRevX.12.031017}.
Autoencoders are used for superconducting qubits by pretraining on qubit readout signals to extract relevant features, which are then used for enhanced classification performance in a supervised manner \cite{PhysRevApplied.20.014045}.
For direct qubit state classification, convolutional neural networks (CNNs) and multilayer perceptrons (MLPs) have been proposed due to their simplicity and ability to mitigate cross-talk effects in multi-qubit configurations \cite{PhysRevApplied.17.014024}, \cite{10.1063/5.0065011}.

In line with previous comparisons between NN models~\cite{PhysRevApplied.17.014024, 10.1063/5.0065011}, we found that different types of neural network architectures perform similarly for the qubit classification problem, with MLPs having a slight edge in classification accuracy.
Beyond accuracy, the choice of NN architecture is highly constrained by the system requirements, available FPGA resources, and strict latency constraints.
 Co-design of the algorithm is a Pareto optimization between algorithm fidelity and NN resources and latency.
Dense MLPs, due to their straightforward design, often require fewer computational resources than more complex architectures like RNNs, CNNs, or autoencoders. 
Considering these factors, we selected an MLP for binary classification to distinguish between the ground and excited states of qubits. 
The model's simplicity ensures effective integration into the QICK system without overwhelming computational resources. 

\begin{figure}[tbh!]
    \centering\includegraphics[width=1\linewidth]{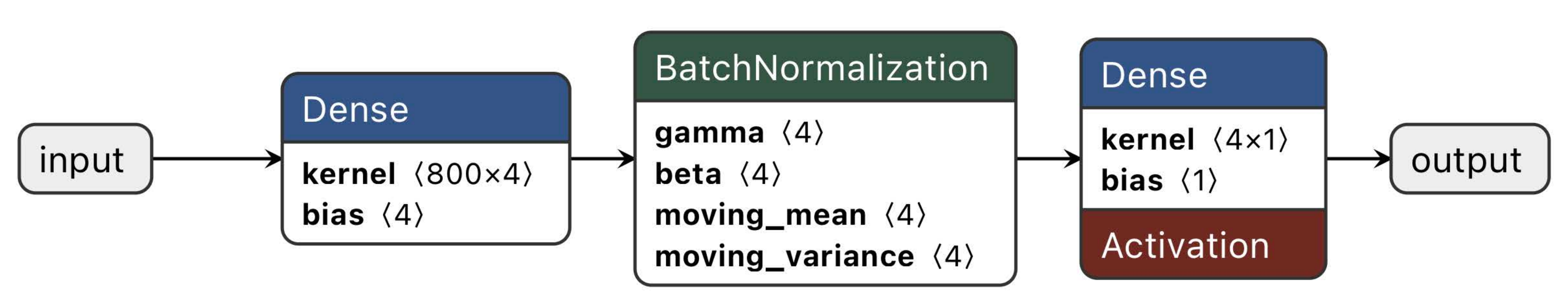}
    \caption{Visualization of a two-layer neural network architecture with an 800-dimensional input, a dense layer with batch normalization followed by another dense layer with a sigmoid activation function.}
    \label{fig:modelarch}
\end{figure}

To that end, we start with a simple 2-layer MLP neural network architecture as illustrated in Fig.~\ref{fig:modelarch}.
We have explored other MLP and CNN architectures, including more and less hidden layers, and we find a 2-layer architecture is sufficient for this task.  The optimal architecture may be different for other tasks such as different experimental setups or multiple qubits. To demonstrate our general co-design principles to minimize FPGA resources while maximizing fidelity performance, we focus on the following:
\begin{itemize}[noitemsep]
    \item Hyperparameter design space exploration: Reducing the overall network size, i.e. weights and computations, through hyperparameter design space exploration -- this includes optimizing the readout window start time and size and the number of neurons in the hidden layers.
    \item Hardware optimization: \textit{Quantization} of the NN through quantization-aware training (QAT) methods -- the resources of the NN approximately scale quadratically with the precision of the operations and embedding quantization into the training process with QAT often yields an overall lower precision model.  Further hardware optimization can be done at the implementation level by tuning the amount of parallelization of the model computations in hardware.  For such low-latency applications, we generally try to ``unroll'' (parallelize) the computations as much as is allowed while balancing FPGA resources.   
\end{itemize}

The model consists of dense and batch normalization layers. 
A batch normalization layer is valuable for such an architecture given the bit width of inputs and the use of fixed-point calculations in quantized neural networks to optimize FPGA resources -- described in more detail below.
Among other benefits of batch normalization, scaling the logits prevents computational overflows.  
We utilized the Adam optimizer \cite{kingma2014adam} and a binary cross-entropy loss function for optimization, and processed the network's output with a sigmoid function to produce a probability distribution over the classes.
We employed gradient descent with a learning rate of \( 10^{-4} \).
Each training iteration was completed in approximately 29 seconds on NVIDIA A100 GPU. 

To design the NN classifier, we adopt \hlsfml~\cite{fastml_hls4ml,Duarte:2018ite}, an open-source software framework that bridges the gap between high-level machine learning models and low-level hardware implementation.
\hlsfml converts machine learning algorithms, especially neural networks, from frameworks like TensorFlow or PyTorch into hardware descriptions in C++ for high-level synthesis (HLS) tools such as AMD/Xilinx Vivado HLS~\cite{vivado_hls}. 
In particular, we will use the QKeras~\cite{qkeras} front-end that interfaces with \hlsfml in order to perform QAT. 
The \hlsfml tool generates a dataflow architecture on FPGA, which is well-suited for neural network computations' parallel and pipelined nature.
In this architecture, each neural network layer can be implemented as a separate hardware module, and data moves sequentially from one module (layer) to the next with minimum buffering.
This allows for continuous data processing and minimizes latency.

\subsection{Hyperparameter Design Space Exploration} \label{subsec:design_space}

\begin{figure}
\centering
\includegraphics[width=1\columnwidth]{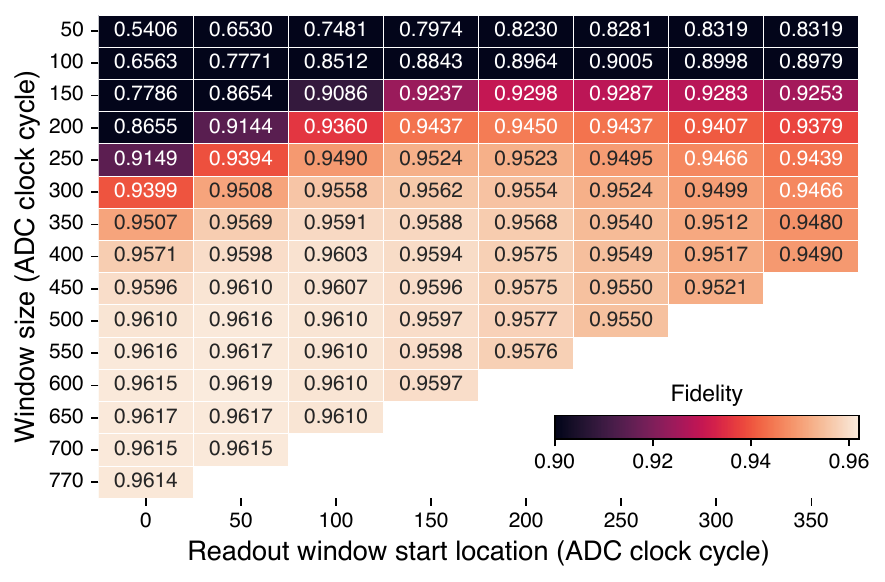}
\caption{Analysis of NN with four hidden neurons performance as a function of the starting time of the readout window (in clock cycles) on the $x$ axis and the size of the readout window on the $y$ axis. The $z$ axis is the fidelity of the model. }
\label{fig:dse_window_size_vs_start_location}
\end{figure}

Figure~\ref{fig:dse_window_size_vs_start_location} shows the fidelity of an optimized neural network with four hidden neurons (hn) using varying readout window sizes and starting locations. 
The smallest window size is 50 clock cycles (CLK) and increments by 50 CLKs up to 700 CLKs, followed by an evaluation on the full readout window. 
This process is repeated for starting locations, beginning at 0 CLK and increasing in 50 CLK increments up to 350.
The results indicate that starting the readout window later generally improves performance, especially with smaller window sizes. 
However, this improvement plateaus at around 100-150 CLKs. Similarly, larger windows lead to better performance, though the benefits diminish beyond approximately 400 CLKs. 
Based on these findings, we focus on a starting location of 100 CLKs, as it provides high fidelity with the smallest effective window size (400 CLKs). 
We repeated this design space exploration for the second NN version with 64 hidden neurons, as well as for thresholding and matched filtering methods, observing similar trends as shown in Figure~\ref{fig:dse_window_size_vs_start_location}.

\begin{figure}
    \includegraphics[width=.45\textwidth]{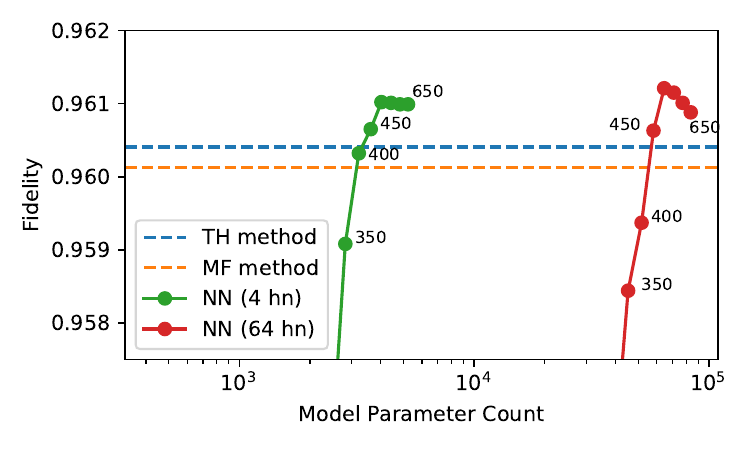}
    \caption{Model size (in parameter count) versus fidelity.  Each point on the graph corresponds to a different window size ranging from 350 to 650 clock cycles.  }
    \label{fig:dse_params_vs_acc}
\end{figure}

Figure~\ref{fig:dse_params_vs_acc} presents the design space exploration of both 2-layer NN variants, comparing their performance based on the number of trainable parameters and test fidelity across different readout window sizes.
All results are based on data starting at 100 CLKs. The exploration includes window sizes starting from 350 CLKs and increasing in 50 CLK increments, up to 650 CLKs.
As seen in Figure~\ref{fig:dse_window_size_vs_start_location}, the most critical features for state discrimination appear after the first 100 ADC clock cycles. 
Therefore, we exclude the first 100 ADC units in our exploration. 
The graph indicates that test accuracy improves with the number of parameters, reaching a peak at a readout window size of 400 CLKs, beyond which additional parameters provide minimal improvement. 
Both NN variants show similar performance within this range, demonstrating that smaller NNs can achieve the same fidelity as larger architectures, making them more efficient.
This may suggest that the smaller NN is less prone to overparameterization and overfitting, making it a more efficient choice for state discrimination and FPGA integration.

\begin{figure}
\centering
\includegraphics[width=.85\columnwidth]{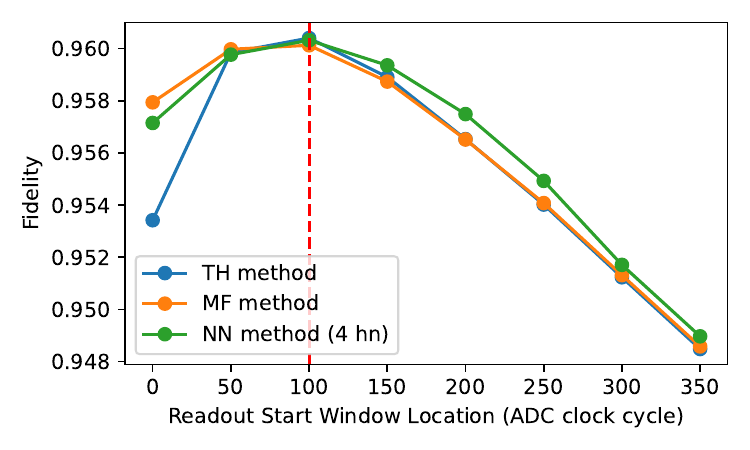}
\caption{Optimal selected neural network hyperparameters (red line) with 400 clock cycle readout window starting at 100 clock cylces compared against the threshold and match filter methods.
}
\label{fig:exp_3}
\end{figure}

Figure~\ref{fig:exp_3} illustrates how fidelity varies as we change the starting position of the 400-CLK readout window of the 2-layer NN with 4 hidden neurons, thresholding, and match filtering methods. 
All methods show a similar trend: fidelity initially increases with the readout start window location, peaks around the 100–150 ADC clock cycle mark, and then gradually declines.
This is consistent with the time it takes to populate the readout resonator, seen previously in the average trajectory data in Fig.~\ref{fig:exp_2}.
When the readout window starts too late, the decrease in fidelity can be associated with qubit $T_1$ decay. 
The NN method generally has a slightly higher fidelity than the TH and MF methods across most locations.
Both methods achieve the highest fidelity when the readout window begins around 100 ADC CLKs. 
Moving forward, we will focus on a 400 CLK readout window starting at 100 CLKs.

\subsection{Hardware optimization}

A co-design approach is taken to optimize neural networks for FPGA implementation. 
Typically, all parameters and data are represented using fixed-point precision, as opposed to the standard 32-bit floating-point notation used during the training and evaluation phases. 
For FPGA deployment, all data is quantized to fixed-point, which results in some level of performance degradation compared to floating-point notation. 
To address this, we employ quantization aware training (QAT). 
This approach helps improve the accuracy of the quantized models on FPGAs by simulating quantization during the training process \cite{jianfei1, jianfei2}. 
Neural networks learn to operate within the constraints of lower precision (e.g., 8-bit integers). 
The benefits of QAT are two-fold: it minimizes the accuracy loss due to quantization and significantly reduces model size, making it easier to deploy on devices with limited memory and hardware resources. 
Quantized models can achieve faster inference times due to the reduced computational complexity of arithmetic operations, which are crucial for real-time processing applications. 
All parameters and activations are quantized using uniform symmetric quantization. 
Data from the ADC units are fed directly to the network as 14-bit unsigned integers. 

In future work, normalizing the ADC data should be studied for robustness against drifts in the readout signal. 
We explore different quantization schemes to evaluate the impact of precision on fidelity. 
The two NN models we tested have $800\times 64\times 1$ and $800\times 4\times 1$ parameters, and we explore their performance when using full precision (32-bit floating-point) and reduced precision, including 6-bit fixed-point, 3-bit fixed-point, and ternary (2-bit) representations. 
Both models show stable performance using 32-bit floating point (32FP). 
6-bit and 3-bit quantization shows a slight drop compared to 32FP, but remains close. 
Ternary quantization performance, 
restricted to $-1$, $0$, or $1$, is similar to 6-bit and 3-bit performance for this single quibit classification task.
All quantization schemes followed the same pattern as 32FP, peaking at 100 CLKs around 96\% fidelity, then declining at the same rate. 
We opt to proceed with the $800\times 4\times 1$ NN using ternary weights to minimize both hardware footprint and latency. 

With a resource-optimized neural network implementation, we can fully parallelize the hardware implementation by unrolling the matrix multiplications. 
This enables us to minimize the computational latency of the ternary neural network.  
We can tune the FPGA resources of the neural network in \hlsfml by configuring this parallelization factor, but we choose to fully unroll the hardware implementation for the lowest latency.  
After synthesis, we find the hardware implementation takes 10 clock cycles in total -- 8 clock cycles for the computation itself and 2 clock cycles to store the results in BRAM.  
With a 3.22\,ns clock cycle, the total latency is 32\,ns.  The 8 clock cycles is driven by the multiplication latency and the number of hidden layers in the neural network.  

\section{Model integration: QICK+\MakeLowercase{hls4ml}}
\label{sec:integration}
\subsection{QICK firmware}
The QICK system adopts a software-based approach in which users can access the system remotely using Jupyter notebooks.
QICK includes, in addition to software applications, \emph{Processing System} (PS) and \emph{Programmable Logic} (PL), as shown in Fig.~\ref{fig:exp-hardware-final}.
The PS in the AMD/Xilinx UltraScale+ device integrates a Zynq system and DDR4 memory, while Linux OS runs on the multicore ARM processor.
QICK uses PYNQ software libraries and drivers to simplify the software-firmware interaction and, in particular, provides the PL with an easy solution for direct memory access (DMA).
The firmware on the PL integrates \emph{Readout} and \emph{Signal Generator} blocks, which are controlled by the timed processor (\emph{tProc}). 
The \emph{tProc} implements custom instructions to produce, for example, pulses to control and readout qubits via the Signal Generator blocks.
In QICK, data flows between firmware components and software applications through the \emph{AXI Interconnect}, which is also the backbone for integrating our NN classifier IP.

\subsection{Neural network classifier integration}
Our NN classifier IP follows the loosely coupled model for hardware accelerators~\cite{cota2015}.
A tightly coupled accelerator would be designed and integrated closely with the \emph{tProc}, increasing its complexity, sharing its caches, and possibly stalling the computation.
Meanwhile, loosely coupled accelerators can be designed separately, easily maintained, and independently integrated. They operate on larger data sets and alternate coarse-grain computation with data transfer phases.
In QICK, we integrated the IP as a device managed with Linux device drivers running on the processing system's ARM cores.
To interface the classifier with the rest of the QICK system, we encapsulated the \hlsfml-generated NN implementation (\texttt{NN\_hls4ml}) in a top-level wrapper (\texttt{NN\_axi}) synthesized in Vivado HLS and described below.

\begin{figure}[h]
\centering
\includegraphics[width=1.\columnwidth]{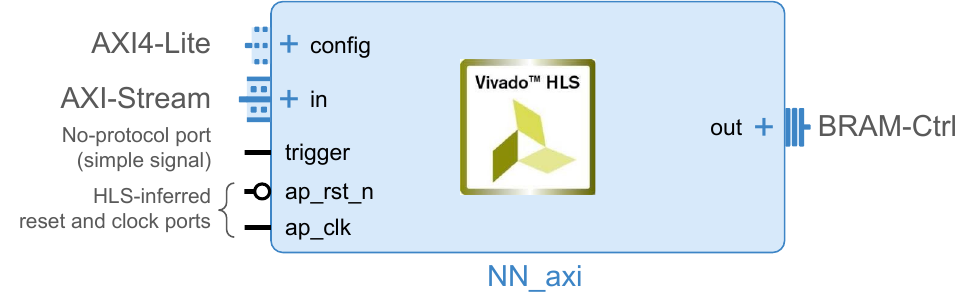}
\caption{The interface of the NN classifier as a Vivado IP integrated into the QICK firmware. AXI interfaces configure and stream readout data; a BRAM interface connects the IP with an external buffer on the programmable logic.}
\label{fig:vivado_hls_ip}
\end{figure}

\begin{figure*}[tbh!]
\centering
\includegraphics[width=.95\textwidth]{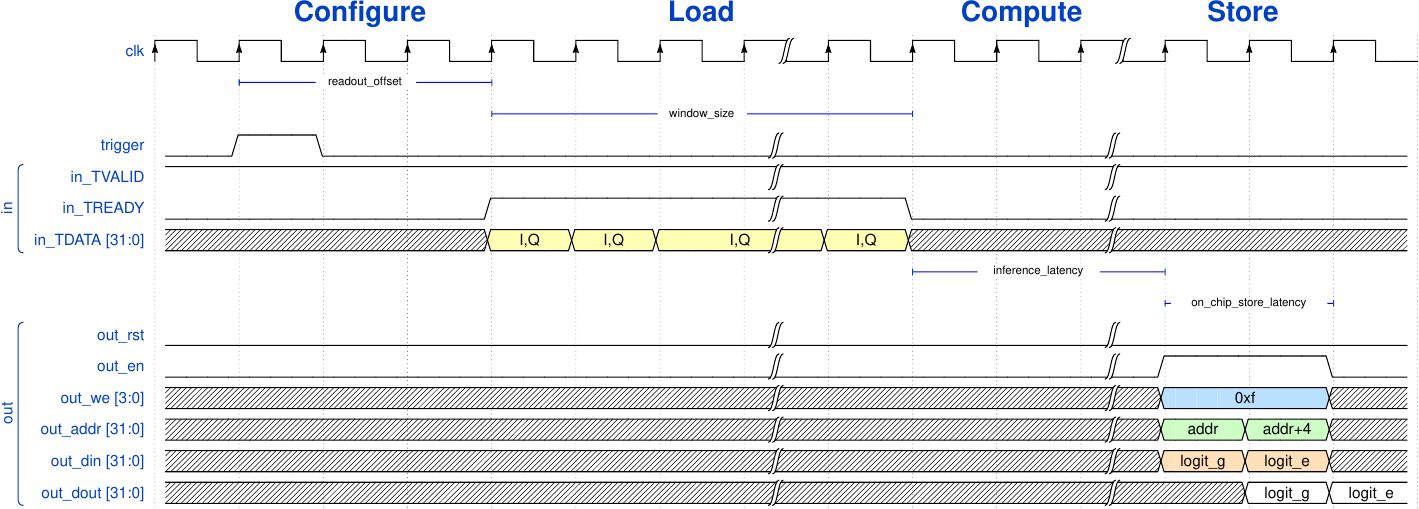}
\caption{Waveforms at the classifier IP interface for a single readout.}
\label{fig:waveforms}
\end{figure*}

\subsection{IP interface}
Figure \ref{fig:vivado_hls_ip} shows the interface of the top-level wrapper (\texttt{NN\_axi}) and NN classifier synthesized with AMD/Xilinx Vivado HLS as an IP for the integration in the QICK Vivado project. 
Four main ports are explicitly defined in the wrapper specifications for HLS:
\begin{itemize}
    \item Port \texttt{config} receives the configuration information from the software via memory-mapped registers, which are bundled together in the AXI4-Lite protocol.
    For example, the user can configure the readout window offset and scaling factor at runtime.
    \item Port \texttt{in} receives input data from the QICK \emph{Readout} block as a discrete-time series of quadrature values (${I_n, Q_n}$).
    The series is defined as a 32-bit wide stream and uses the AXI-Stream (AXIS) protocol with no \texttt{TREADY} or \texttt{TLAST} side-channel signals.
    Each 32-bit word packs two 14-bit I and Q values with four bits of zero padding.
    Typically, with an AXIS protocol, the \texttt{TREADY} signal allows the receiver to control the pace of the data transfer, indicating to the transmitter when it is ready to accept more data (i.e., backpressure).
    In QICK, the lack of backpressure towards the \emph{Readout} block means that the receiver must always be ready to accept data. 
    \item Port \texttt{out} transfers the classifier output logit values to a memory buffer on the programmable logic block memories (BRAM).
    The buffer size is 128KB, which allows the storage of 16,384 consecutive predictions.
    The BRAMs are mapped to the main memory via AXI4-Lite and function as high-speed, local memory within the PL, allowing the NN to save the prediction in a buffer with minimum delay.
    The AXI4-Lite interface provides a communication link between the software running on the RFSoC processor and the BRAMs, enabling access to the prediction results at a lower throughput.
    The BRAMs on the PL and their corresponding regions in the main memory operate independently, with the PL logic directly managing the data in the BRAMs, while the AXI4-Lite interface allows the processor to interact with this data in a controlled manner following a loosely coupled paradigm.
    \item Port \texttt{trigger} receives the trigger signal from the \emph{tProc} that controls the classifier execution.
    No HLS-generated protocol is specified for this port, and the associated signal is used directly in the control logic of the IP.
\end{itemize}
Vivado HLS automatically infers additional ports for clock and reset signals and adds them to the final IP interface, as shown in Figure \ref{fig:vivado_hls_ip}.

\subsection{Behavior and Timing}

The classifier IP execution has four phases: configuration, data loading, NN computation, and storage of the prediction results.
First, the software application configures the accelerator via memory-mapped registers.
Readout data is then streamed over an AXI-stream interface during the loading phase and transferred to the IP's private local memory.
The NN prediction results are saved locally to BRAM in the PL and finally transferred to the system's main memory through an AXI4-Lite interface.

Figure \ref{fig:waveforms} shows the behavior of the classifier as waveforms at the IP interface for a single readout trace.
The clock period is 3.25~ns.
After a pulse of the \texttt{trigger} signal, the classifier loads the input data from the \texttt{in} channel, which has a data lane (\texttt{in\_TDATA}) that is 32-bits wide to pack two 14-bit I and Q values.
Data constantly stream from the ADC to the readout block and into the classifier; thus the valid signal (\texttt{in\_TVALID}) is mostly high.
Since the classifier IP does not push back, the ready signal (\texttt{in\_TREADY}) is only shown as a reference.
The user can configure an additional delay (the \texttt{readout\_offset}) counted in clock cycles from the rising edge of the trigger pulse. 
The duration of the load phase in clock cycles is the \texttt{window\_size} of the trained neural network.
The \texttt{inference\_latency} depends on the complexity of the neural network, and varies between 5 and 20 clock cycles for the different NN models we tested.
Finally, the IP takes two more clock cycles (\texttt{on\_chip\_store\_latency}) to store the inference results locally in the programmable logic in BRAMs through a simple memory interface \texttt{out}.

\subsection{Software API and Usage}

We defined a comprehensive Python API that runs on the RFSoC processor and interacts with the NN classifier integrated with the rest of QICK system on the programmable logic.
These functions enable users to reset, configure, and retrieve predictions from the classifier, as well as debug and inspect the classifier's internal state.

Once the bitstream is loaded on the FPGA, the first function to be called is \texttt{reset\_classifier}.
This function ensures that the classifier is in a clean state by resetting its configuration and state registers.
Depending on the needs of the application, a deep reset may be performed to zero out specific memory locations, providing a fresh start for subsequent operations. 
The parameters \texttt{index\_lo} and \texttt{index\_hi} allow for selective resetting of a range of memory indices.

After resetting, the classifier is configured using the \texttt{configure\_classifier} function.
This step involves setting parameters such as the \texttt{readout\_offset} or the \texttt{scaling\_factor}, which define the operational characteristics of the classifier.
Additionally, the \texttt{debug} parameter can be set to true to enable detailed logging of the configuration process, aiding in debugging and fine-tuning.

\begin{table*}[th]
\centering
\begin{tabular}{@{}lcccc@{}}
\toprule
                       & \multicolumn{2}{c}{\textsc{Memory Resources}} & \multicolumn{2}{c}{\textsc{Computational Resources}} \\
                       & \textsc{FF} & \textsc{BRAM} & \textsc{LUT} & \textsc{DSP} \\
\midrule
\textsc{ZCU216}        & 850,560     & 1,080        & 425,280      & 4,272       \\
\textsc{QICK}          & 89,783 (10.56\%) & 309 (28.61\%) & 60,057 (14.12\%) & 481 (11.26\%) \\
\textsc{QICK+NN}       & 124,152 (14.60\%) & 341 (31.57\%) & 126,726 (29.80\%) & 481 (11.26\%) \\
\textsc{NN}            & +4.04\% & +2.96\% & +15.60\% & +0.00\% \\
\bottomrule
\end{tabular}
\caption{Overall FPGA resources available on Zynq UltraScale+ RFSoC ZCU216 and resource utilization for QICK and its ML-enhanced version. The memory resources are flip-flops (FF) and block RAMs (BRAM); the computational resources are look-up tables (LUT) and data-signal processors (DSP).\label{tab:res_utilize}}
\end{table*}

After the initial state preparation, we use a QICK program that runs on the \emph{tProc} and sends a square-shaped readout pulse to the readout resonator.  
The ADC digitizes the returned readout signal to a discrete-time series of quadrature values ${I_n, Q_n}$ that feeds our classifier.
At this point, to monitor the classifier's performance, the \texttt{get\_classifier\_prediction\_count} function returns the total number of predictions made by the classifier.
This count starts at zero upon initialization or after a reset, providing a clear indication of how many pulses have been sent and predictions have been processed.

The \texttt{get\_classifier\_prediction} functions allow users to retrieve the classifier's output for a specific index or a range of indices.
These outputs are returned as tuples containing logits for the ground and excited states, which can be further analyzed or used in subsequent processing stages.

Finally, to inspect the classifier's internal state, the \texttt{print\_classifier\_buffer} function prints the contents of the classifier's buffer for a specified range of indices. 
This is particularly useful for verifying that the classifier has processed the data correctly and can be an essential tool for debugging.

\subsection{Performance and Implementation}

In summary, our final implementation of the NN model performs single-qubit state discrimination with a readout pulse length of 500 clock cycles ($1.6\,\mu$s). Using an ADC offset of 100 clock cycles, the last 400 clock cycles of the readout signal are streamed to the NN block. The NN takes 8 clock cycles to perform the inference and an additional 2 clock cycles to store the inference result in a memory buffer, corresponding to a total latency of 10 clock cycles (32\,ns) following the end of the readout window.

With the NN integrated readout, we performed 500,000 single-shot readout experiments for both the ground- and excited-state readouts.
To directly assess the NN discriminator's performance, we record both the NN predicted outputs and the raw time-series I/Q data for each single-shot experiment. 
The NN state discriminator resulted in a readout fidelity of $\mathcal{F}=96$\%, comparable to the performance of the simple thresholding or matched filter methods over the same readout window. For a single-qubit system, we expect similar performance and will now be able to deploy this end-to-end flow in more complex and dynamic systems -- discussed in more detail in Section~\ref{sec:summary}.

Table~\ref{tab:res_utilize} shows the resource requirements for implementing the model in FPGA. Over the columns, we report the memory resources as flip-flops (FF), basic memory elements used to store binary data, and Block RAMs, dedicated memory blocks that can store up to 36 kilobits on the programmable logic; the computational resources are look-up tables (LUT), configurable logic blocks used for implementing logic functions, and digital signal processors (DSPs), specialized hardware units for efficient computation of operations like multiplications and additions.
The row denoted by \textsc{ZCU216} shows the overall resources available on the chip of our target development board (Zynq UltraScale+ RFSoC ZCU216). The second row shows the resources required by the \textsc{QICK} platform, while the third row details the combined resources for the QICK plus the NN IP; in both cases, we report the absolute value and the percentage of usage for each resource (in parentheses). Finally, the last row indicates the overhead of the NN IP alone. 
The NN IP requires additional resources of 4.04\% for FFs, 2.96\% for BRAMs, and 15.6\% for LUTs.
Finally, in terms of performance, once the readout data has been loaded, the algorithm latency is ten clock cycles: eight cycles for the NN inference and two cycles to store the results in the external BRAMs.

\section{Summary and Outlook}
\label{sec:summary}

This study introduces a comprehensive workflow for enhancing the readout of superconducting qubits by incorporating co-designed NN into the QICK hardware platform. Using \hlsfml to efficiently co-design NNs on programmable logic, the workflow addresses critical challenges such as improving accuracy, reducing latency, and preserving quantum states during readout. The NN algorithm is optimized using ternary quantization and parallelization methods to run with a low latency of 32\,ns following the qubit readout process, consuming less than 16\% of the FPGA look-up table resources and less for other resources. Performance evaluations show that this approach achieves fidelity comparable to conventional techniques like thresholding and matched filtering for single qubit readout. This open-source framework demonstrates the feasibility of NN-based readout for superconducting qubits. It provides a valuable tool for researchers to explore innovative quantum computing methodologies that integrate machine learning with high-level synthesis for efficient hardware deployment.

There are increasingly more studies on embedded, real-time, ML-based qubit readout such as in Ref~\cite{lienhard2022deep,vora2024ml,gautam2024low}. However, the experiments from which the data come and, thus, the algorithmic approaches, optimization, and implementation vary.  To make direct comparisons of different approaches and new methods as they are developed, it is valuable, as a community, to have publicly available benchmarks, including common datasets and reproducible results.  To that end, we have made our dataset available on Zenodo~\cite{zenodo-ml-quantum-readout} and the code to reproduce the algorithm training and implementation available at Ref~\cite{github-ml-quantum-readout}.  We hope that this will be useful for future studies and encourage more datasets and benchmarks to be made available as systems grow in complexity.  

One near-term upcoming for our platform for user experimentation is to integrate output of the NN block into the conditional logic of QICK to run readout experiments including real-time ML feedback control. 
This will also be updated and included in our publicly available code repository.

There are several directions of exploration that follow-on directly from this work towards realizing the ultimate goal of an adaptive and continuously optimized readout system.  
For example, model-based readout~\cite{PhysRevLett.132.100603} and reinforcement learning methods~\cite{chatterjee2024demonstration} are promising for continuous and autonomous qubit readout.  This could be integrated directly with the QICK and \hlsfml platforms. While our demonstration algorithm uses a straightforward dense neural network architecture, there is significant potential for developing more performant and resource-optimized algorithms in hardware using additional codesign methods such as pruning~\cite{cheng2024survey} or other efficient architectures amenable to time-series data~\cite{gu2022train}.  Relatedly, developing robust algorithms to changing instrument conditions~\cite{baldi2024reliable} can also aid in improved continuous learning.  These directions of exploration are especially valuable as readout systems grow in complexity for multi-qubit systems, and ML approaches are already proving to be powerful~\cite{gautam2024low,Duan2021-zg,PhysRevLett.114.200501,Maurya2023scaling}.

In the longer term, we plan to deploy the control and readout logic in the cryostat as a more scalable solution for quantum computers with many thousands of qubits.
Placing an SoC that integrates programmable logic as embedded FPGAs in a cryostat for quantum readout is motivated by the need to minimize thermal noise, enhance signal integrity, and reduce latency.
The close proximity of the logic to the qubits within the cryostat minimizes signal loss and noise introduction, leading to a more accurate and reliable quantum readout.
Additionally, the reduced latency in signal processing is essential for real-time quantum error correction and feedback, where even minor delays can impact system performance.
Integrating classical control hardware with quantum hardware in the same cryogenic environment also simplifies system design, improves efficiency, and supports scalability as quantum computing systems become more complex~\cite{chakraborty2022cryo,frank2022cryo}.

\section*{Acknowledgements}


We would like to thank Sho Uemura, Leandro Stefanazzi, and Gustavo Cancelo of the QICK development team for their technical assistance.

This document was prepared using the resources of the Fermi National Accelerator Laboratory (Fermilab), a U.S. Department of Energy, Office of Science, Office of High Energy Physics HEP User Facility. Fermilab is managed by Fermi Forward Discovery Group, LLC, acting under Contract No. 89243024CSC000002.

G.N.P. was supported for this work by the Department of Energy (DOE), Office of Science, Office of High Energy Physics, QuantISED program grant ``HEP Machine Learning and Optimization Go Quantum,'' identification number 0000240323.
N.T. and J.C. are supported by the U.S. Department of Energy (DOE), Office of Science, Office of Advanced Scientific Computing Research under the “Real-time Data Reduction Codesign at the Extreme Edge for Science” Project (DE-FOA-0002501).
G.D.G. is supported by Fermi Research Alliance, LLC under Contract No. DE-AC02-07CH11359 with the Department of Energy (DOE), Office of Science, Office of High Energy Physics. B.D., S.L., R.M., and D.B. are supported by the U.S. Department of Energy, Office of Science, National Quantum Information Science Research Centers, Quantum Science Center. 

\bibliography{references}

\end{document}